\documentclass[unnumbib]{comnet}

\usepackage{graphicx}
\usepackage{amsmath}
\usepackage{xcolor}
\usepackage{natbib}
\setcitestyle{numbers}
\usepackage{multirow}
\usepackage{array}

\begin{document}

\title{Effects of Time Horizons on Influence Maximization in the Voter Dynamics}

\shorttitle{Effects of Time Horizons on Influence Maximization in the Voter Dynamics} 
\shortauthorlist{Brede et al.} 

\author{
\name{Markus Brede$^*$}
\address{ECS, University of Southampton\email{$^*$brede.markus@gmail.com}}
\name{Valerio Restocchi}
\address{ECS, University of Southampton}
\and
\name{Sebastian Stein}
\address{ECS, University of Southampton}}

\maketitle

\begin{abstract}
{In this paper we analyze influence maximization in the voter model with an active strategic and a passive influencing party in non-stationary settings. We thus explore the dependence of optimal influence allocation on the time horizons of the strategic influencer. We find that on undirected heterogeneous networks, for short time horizons, influence is maximized when targeting low-degree nodes, while for long time horizons influence maximization is achieved when controlling hub nodes. Furthermore, we show that for short and intermediate time scales influence maximization can exploit knowledge of (transient) opinion configurations. More in detail, we find two rules. First, nodes with states differing from the strategic influencer's goal should be targeted. Second, if only few nodes are initially aligned with the strategic influencer, nodes subject to opposing influence should be avoided, but when many nodes are aligned, an optimal influencer should shadow opposing influence.}
{Insert keyword text here.}
\\
2000 Math Subject Classification: 34K30, 35K57, 35Q80,  92D25
\end{abstract}


\section{Introduction}
	Influence maximization --the study of strategically influencing agents on social networks with the aim to align their opinions or choices with certain targets-- has major applications ranging from advertising and marketing to the political campaign problem \citep{Hegselmann:2015}, analyzing the spread of extreme opinions and radicalization \cite{Galam2016:2016}, or limiting the spread of fake news \citep{Nguyen:2012}. The underlying diffusion process has been extensively studied in competitive and non-competitive scenarios \citep{Kempe:2003}, mostly via variants of models based on the seminal independent cascade model \citep{Bharathi:2007,Borodin:2010,Goyal:2014,Morone:2015}. In the independent cascade model agents are committed to an opinion once affected by contagion, but are not subject to dynamic forces of opinion change thereafter. Such a scenario may be appropriate when there is an effect of lock-in, e.g., if agents make one-off decisions about buying an expensive product or generally making decisions when further change is costly. However, as it has been realized by some authors \citep{Yildiz:2013,Kuhlman:2013,Masuda:2015,Brede:2018}, these models may not be appropriate in situations in which agents are subject to various sources of social influence, and decisions can be changed over time. Such decision making can be described by various types of dynamic models of opinion formation for which there is a rich interdisciplinary literature (see, e.g., \citep{Castellano:2009} or \citep{Sirbu:2016} for recent reviews). One can generally distinguish between models that either consider opinions as discrete (i.e. as a stance for or against a certain issue under discussion) or as continuous, expressing gradations of closeness in opinion spaces. Whereas most modeling effort has been devoted to understanding various facets of opinion dynamics, some recent studies have also started to consider influence maximization for such dynamic models of opinion formation. In this context, previous work has focused on exploring optimal control allocations that maximize influence in the stationary state of the kinetic Ising model \citep{Liu:2010,Laciana:2011,Lynn:2016}, for the AB model \citep{Arendt:2015} and for the voter dynamics \citep{Yildiz:2013,Kuhlman:2013,Masuda:2015,Brede:2018}. 
    
    Here, because of its rich history, we concentrate on the voter dynamics \citep{Clifford:1973,Holley:1975} which describes the dynamic choice between two discrete opinions. Importantly, compared to a contagion model in which adopting a state is a one-off process, in the voter dynamics agents repeatedly update their opinion states depending on the state of their social network neighbors, which allows to study influence maximization in a dynamic scenario in which agent states can switch back and forth. Many exact results about the voter dynamics are known \citep{Castellano:2009}, showing that without external influence a consensus will be reached on finite networks and infinite low-dimensional lattices \cite{Castellano:2009}. External influence to the voting dynamics, often also called zealotry, is typically modeled by agents with a bias \citep{Mobilia:2003,Mobilia:2005} or unidirectional influence \citep{Mobilia:2007} and can dramatically alter this picture and lead to equilibrium states in which multiple opinions can coexist. Zealotry in the voting dynamics has been extensively studied \cite{Mobilia:2003,Mobilia:2005,Mobilia:2007}, with most recent extensions to linear and non-linear multi-state voter models \citep{Javarone:2015,Waagen:2015,Mobilia:2015,Hu:2017} or studies on the effects of zealots in the noisy voter model \citep{Khalil:2018}. 
    
    Recent work \citep{Yildiz:2013,Kuhlman:2013,Masuda:2015,Brede:2018} has also started to address the question of opinion control in the voter model. Up to our best knowledge, all work in the area has focused on maximizing vote shares in the stationary state, finding that albeit centrality measures cannot exactly predict optimal allocations \citep{Yildiz:2013} optimal control will generally focus on high-degree nodes on undirected networks \cite{Kuhlman:2013,Masuda:2015}, at least if the resistance of nodes to control is small \cite{Brede:2018}. However, the transient dynamics of the voting process can sometimes be long and influence maximization in real-world settings may be targeted at achieving best results in shorter time frames \citep{Hegselmann:2015}. Even though previous work provides some approximations for transient times in the voter model \citep{Yildiz:2013}, the question of non-stationary influence maximization-- the topic of this paper-- has not found much attention for the voter dynamics. Up to our best knowledge, the only study investigating effects related to non-stationary influence maximization in a voter-like model is \cite{Javarone:2014} in which the author compares the effectiveness of various "local" and "global" influence allocation strategies. This study, however, does not relate the effectiveness of such strategies to equilibration times and time horizons of planners in the voting dynamics.

Further, closest to our paper, recent work of Alshamsi et al. \citep{Alshamsi:2017} has investigated the effects of time scales in independent-cascade-like models of strategic diffusion when considering complex contagion. As opposed to simple contagion in which infection of a single neighbor is sufficient to infect a new node (as, e.g., when considering epidemic spreading \citep{Pastor:2015}), empirical studies have suggested that in opinion formation reliable contagion requires reinforcement of exposure through multiple sources \citep{Centola:2007,Centola:2010}. Alshamsi et al. show that influence maximizers should target nodes of different degree at different stages of the contagion process, i.e., starting with low-degree nodes and then hub nodes only when the cascade has propagated beyond a certain point. However, the model of Alshamsi et al. is not a dynamic model in the sense of the voter model, i.e., it is not appropriate for situations of fast opinion switching and does not allow to study the dynamic balance between different sources of influence. This is however fundamental to model to mimic real-world contexts such as marketing \citep{Mehmood:2016} or governmental intervention strategies (e.g., public health campaigns \citep{Yadav:2016} or radicalization prevention \citep{Ramos:2015}), where two or more parties compete with each other to influence people towards the adoption a specific opinion. Examples of this scenario are plenty, e.g., two firms with the same target audience which want to convince people to adopt their product over their competitors, or a public campaign in which the government's aim is to incentivise people to make healthy lifestyle choices instead of drinking or eating junk food, which would clash with the influence that companies selling these products would try to exert on people to become their customers. Moreover, in such scenarios, the time horizon is often fixed and potentially very short, and analyzing the effect of such a constraint on influence maximization is paramount to design algorithms that efficiently tackle this problem.

Our results suggest that optimal influencing strategies differ depending on the time horizons of the strategic influencer. Specifically, we find that low-degree nodes represent better control targets when the time horizon is short, whereas hubs are ideal targets for long time horizons. Moreover, our results suggest that optimal strategies vary depending on the initial state of the network. Importantly, we find that, when many nodes are aligned with the strategic influencer's opinion, its best strategy is to neutralize its opponent's influence. Conversely, when most voters hold a different opinion than the strategic influencer, the best strategy is to directly target opposing low-degree nodes and avoid the opponent's influence.

The remainder of the paper is organized as follows. In Sec. \ref{Model} we briefly revisit the voter dynamics on complex networks and describe the framework we use for influence maximization. In Sec. \ref{Results} we outline our main results, first, by giving analytical arguments for star networks in subsection \ref{Star} and developing a mean-field theory for degree heterogeneous networks in subsection \ref{MF}.  We then report results from numerical optimization in \ref{sec:2} and, based on these findings, develop static and dynamic heuristics which are explored in subsections \ref{sec:3} and \ref{sec:4}. The paper concludes with a summary and discussion.

\section{Model}
\label{Model}
In the following we consider the voter model first introduced in \citep{Clifford:1973,Holley:1975} on a social network. Each of $N$ nodes of the network is identified with a voter who can be in two possible states $s_i=A$ or $s_i=B$. We assume social networks to be undirected, and thus, their adjacency matrices are given by $G=\{a_{ij}\}_{i,j=1}^N$ with $a_{ij}=1$ if agent $i$ is connected to agent $j$ and $a_{ij}=0$ otherwise. Provided the networks are connected (or strongly connected in the case of directed networks) the voting dynamics is known to reach a consensus on finite networks \citep{Castellano:2009}. Hence, on top of the population of social voters we also consider two influencers -- agents labeled $A$ and $B$ who have directed connections to social voters but are not influenced by them in return. Influencers (or, using the parlance of voter dynamics, zealots) thus have static opinions $s_A=A$ and $s_B=B$ and aim to sway the overall vote of the agent population towards their own opinions.  Below, we will be interested in a scenario in which influencer B is passive with a random allocation of influenced voters. In contrast, we will treat A as active in the sense that we seek configurations of influenced nodes such that A maximizes its influence on the system. We distinguish two types of active allocations, which we term static and dynamic below. In static allocations -- used in the majority of the paper except for subsection \ref{sec:4} -- A decides its influence allocation at time $t=0$ and this allocation then remains fixed. In dynamic allocations, treated in subsection \ref{sec:4}, A changes its influence allocation dynamically over time.

Voters interact with their network neighbors according to the following dynamics: (i) a voter is picked at random, (ii) the voter randomly selects one of its incoming network connections, and (iii) the voter copies the state of the selected in-neighbor. The voter model has been well studied in the literature  (see, e.g. \citep{Castellano:2009} or \citep{Sirbu:2016} for reviews) and analytical solutions for various scenarios are available. Here we follow the approach of Masuda \citep{Masuda:2015}, and use the master equation to calculate occupation probability flows. Let $x_i$ be the probability that agent $i$ is in state $A$. We then have


\begin{equation}
 \label{E0}
 \Delta_i \frac{dx_i}{dt}=(1-x_i) (\sum_j a_{ji} x_j + p_{A,i}) - x_i (\sum_j a_{ji} (1-x_j)+p_{B,i}),
\end{equation}
where $\Delta_i=\sum_j a_{ji} + p_{A,i}+p_{B,i}$ and $p_{A,i}$ and $p_{B,i}$ are the \emph{control gains}, and are set to one if A or B influence $i$, respectively, and zero otherwise. In Eq. (\ref{E0}) the first term corresponds to voters who hold opinion $B$ but are converted to opinion A in the current update; the second term gives voters who hold opinion A but are converted to B due to contact with a voter who holds opinion B. The normalization $1/\Delta_i$ reflects the choice of incoming links at random when updating opinions. Eq. (\ref{E0}) can be simplified to
\begin{equation}
\label{E1}
 \Delta_i \frac{dx_i}{dt}=p_{A,i} -x_i (p_{A,i}+p_{B,i}) + \sum_j a_{ji} x_j - x_i \sum_j a_{ji},
\end{equation}
i.e., we obtain a linear inhomogeneous system of first order differential equations to describe the dynamics of opinion change, where the inhomogeneity is given by the control gains. In previous work, Masuda \citep{Masuda:2015} has studied optimal opinion control by analyzing stationary states of Eq. (\ref{E1}), which can be obtained by solving the corresponding linear system of equations. Here instead we are interested in the transient dynamics of system (\ref{E1}). In principle, this linear inhomogeneous system could be solved exactly via eigenvalue decomposition and, e.g., the method of variation of the constant. We will employ this technique to treat example network configurations below. For numerical results, however, we made a pragmatic choice and solve (\ref{E1}) via numerical integration using a Runge-Kutta method. Step sizes are chosen as $\Delta t=1/N$, so that one integration step could roughly be equated to one discrete update of votes.

To obtain influence-maximizing configurations for influencer A we thus proceed as follows: (i) start with a given social network configuration and randomly chosen initial assignments of votes to nodes (excluding the influencers A and B). To investigate the influence of different initial conditions we set  states of nodes to $s=A$ with probability $q$ and $s=B$ with probability $1-q$. We also start with a random initialization of controls $p_{A,i}$ and $p_{B,i}$ subject to the constraint $\sum_i p_{A,i}=n_A$ and $\sum_i p_{B,i}=n_B$ of given resources $n_A$ and $n_B$ available to the influencers. (ii) To estimate the influence exerted by A we then integrate Eq. (\ref{E1}) over the interval $[0,T]$ and record the average influence of A at time $T$ obtained as $X_A(T,p_A)=1/N\sum_i x_i(T)$. Next, stochastic hill-climbing is used to optimize A's influence, i.e., we randomly pick a voter $i$ controlled by A and rewire to a yet uncontrolled voter $j$ if $X_A(T, p_{A,i}=0,p_{A,j}=1)>X_A(T, p_{A,i}=1,p_{A,j}=0)$. Note that self-connections or multiple connections are prevented when suggesting rewiring. Step (ii) is repeated until no further improvements in $X_A(T)$ have been found for a large number of $N_R$ rewiring attempts. We typically set $N_R=10 N$, such that each node has been tested roughly 10 times before rewiring is aborted.

The focus of our study is on strategies for influence maximization on heterogeneous networks. Accordingly, we will analytically investigate a simple star network configuration \and analyze a degree-based mean-field approximation to gain some intuition about the effects of opinion control targeting hub nodes when time horizons are finite.  We will complement these results with a detailed numerical investigation of scale-free networks with degree distribution $P(k) \propto k^{-\alpha}$ constructed via the configuration model~\citep{Newman:2010}. The choice of this model allows us to investigate dependencies on degree heterogeneity by tuning $\alpha$, noting that, when fixing average degree, networks constructed for $\alpha \gg 1$ become more and more similar to regular random graphs \cite{Bara}. Below, when tuning $\alpha$ we vary proportionality constants to ensure networks with the same average connectivity $\langle k\rangle$ are compared (and prevent multiple and self-connections in the process of network generation).

\section{Results}
\label{Results}
\begin{figure*}[tbp]
 \begin{center}
\includegraphics[width=.95\textwidth]{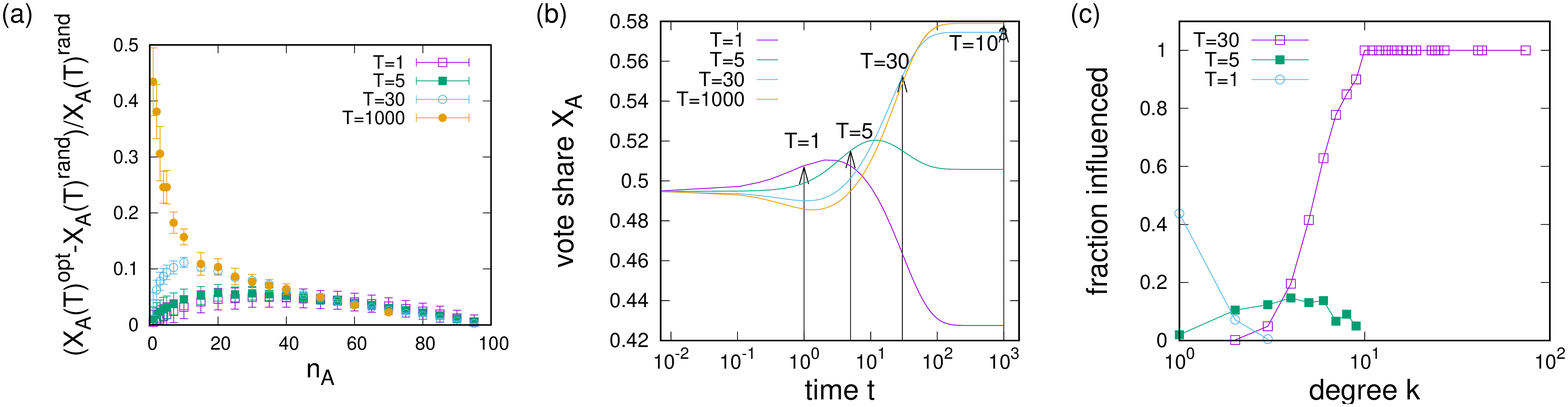}
 \caption {(a) Dependence of optimization gain $(X_A^\text{(opt)}-X_A^\text{(rand)})/X_A^\text{(rand)}$ on resources $n_A$ available to the active voter when competing against a passive voter with resource $n_B=10$ which is randomly allocated. (b) Comparison of time evolution of vote shares for influence maximization schemes evolved for time horizons $T=1$, $T=5$, $T=30$, and $T=1000$ for $n_A=n_B=10$. (c) Dependence of the fraction of influenced nodes for $T=1,5$ and $30$ on node degree $k$. Results are for networks of $N=100$ nodes with $\alpha=3$ averaged over 100 runs for $T\leq 30$ and 30 runs for $T=1000$. Networks have average degree $\langle k\rangle=3$ and initially 50\% of votes have been randomly allocated to both types.}
\label{fig.0}
\end{center}
 \end{figure*}

We start our analysis with an analytical investigation of a star network, a graph topology which is found to represent real-world phenomena such as communication patterns in social media \citep{Chin:2017}. Such an analysis will give insights about the dependency of node controllability on degree for various time horizons T. We then proceed with a detailed numerical investigation for larger scale-free networks in which we demonstrate the robustness of our findings for more complex network topologies and present further results about refined control strategies that can exploit knowledge of initial conditions when time horizons are finite.

\subsection{Analysis of a star network}
\label{Star}
We consider a star network consisting of one hub (labeled 0) with $k$ neighbors (labeled 1). More formally we have $a_{ij}=a_{ji}=1$ for $i=0$ and $j=1,...,k$ and $a_{ij}=0$ for all other combinations of $i$ and $j$. As all the peripheral nodes (or \emph{spokes}) $k=1,...,N$ are topologically identical, we do not need to discriminate between them provided initial conditions are the same, i.e., $x_1(0)=x_2(0)=...=x_k(0)$. We can then rewrite Eq. (\ref{E0}) and obtain
\begin{align}
\label{E2}
 \dot{\vec{x}}=\begin{bmatrix}
           1/(k+1)\\ 0
          \end{bmatrix} +
          \begin{bmatrix}
           -1 & k/(k+1) \\
           1 & -1
          \end{bmatrix} \vec{x},
\end{align}
with $\vec{x}=(x_0,x_1)^T$. The corresponding eigenvalues are
\begin{equation}
\label{eigenvalues}
 \lambda_{1/2}=-1 \pm \sqrt{1-\frac{1}{k+1}},
\end{equation}
which determine the time scales of opinion change in the above system. Both eigenvalues are negative, indicating, as expected, that the stationary state $x_0=x_1=1$ of perfect control is stable for all $k$. However, in leading order in $k$, one has $\lambda_1 \approx -1/(2k)$ and $\lambda_2\approx -2+1/(2k)$ already indicating a slow-down of the convergence towards the stationary state for large $k$. Solving Eq. (\ref{E2}) with initial conditions $x_1(0)=x_2(0)=...=x_k(0)=0$ we can obtain a solution that gives the dynamics of approaching the perfectly controlled state from an initially completely unaligned state. We obtain (for more details, see Appendix A)

 \begin{equation}
 \label{E3}
 x_0(t)=1+\frac{1}{2}e^{-\left(1+\sqrt{\frac{k}{1+k}}\right)t} \left(\sqrt{\frac{k}{1+k}}-1\right) -\frac{1}{2} e^{\left(-1+\sqrt{\frac{k}{1+k}}\right)t} \left(1+\sqrt{\frac{k}{1+k}}\right)
\end{equation}
and
\begin{equation}
\label{E4}
 x_1(t)=1+\frac{1}{2}e^{-\left(1+\sqrt{\frac{k}{1+k}}\right)t}\left(\sqrt{\frac{1+k}{k}}-\frac{1}{k}\right)-\frac{1}{2}e^{\left(-1+\sqrt{\frac{k}{1+k}}\right)t}\left(1+\sqrt{\frac{1+k}{k}}\right)
\end{equation}

  Our main interest is in the dependence of expressions (\ref{E3}) and (\ref{E4}) on $k$ with particular emphasis on hub behavior for large $k$. Expanding in leading order in $k$ we obtain the following asymptotic behavior which is identical for $x_0(t)$ and $x_1(t)$
  \begin{align}
   \label{E5}
   x_{0/1}(t) \approx x_1(t) \approx  1-\exp (-t/(2k)),
  \end{align}
  and thus, since $X_A(t)=(x_0(t)+kx_1(t))/(k+1)$, we obtain $X_A(t) \approx 1-\exp (-t/(2k))$.
  In other words, Eq. (\ref{E5}) shows that hub nodes only become aligned to the target of their influencer at a time scale proportional to their degrees, i.e., successfully influencing a large degree hub node takes much longer than gaining control over a lower degree node. Thus, whereas controlling hub nodes may seem a good strategy in the stationary state \citep{Kuhlman:2013,Masuda:2015}, if the aim is to optimize vote shares for times $t \ll 2k$ it is a better strategy to control lower degree nodes. We will explore this issue in more detail in numerical experiments in subsection \ref{sec:2}.
  
Before proceeding, let us briefly consider a star network which is being influenced by two controllers. System (\ref{E2}) will then have to be modified as follows
\begin{align}
\label{EN0}
 \dot{\vec{x}}=\begin{bmatrix}
           1/(k+2)\\ 0
          \end{bmatrix} +
          \begin{bmatrix}
           -1 & k/(k+2) \\
           1 & -1
          \end{bmatrix} \vec{x}.
\end{align}
It is immediate to note that the stationary state of (\ref{EN0}) now becomes $x_0=x_1=1/2$, instead of $x_0=x_1=1$ as in the previous case of a single influencer, i.e., states will stochastically flip between A and B and, on long time scales, both influencers will be able to exert control over the system to equal amounts. Eigenvalues now become
\begin{equation}
 \lambda_{1/2}=-1 \pm \sqrt{1-\frac{1}{k+2}},
\end{equation}
and thus for $k>1$ the slowest convergence time is related to $T_\text{conv.} \approx k+2$ instead of $T_\text{conv.} \approx k+1$ for the scenario with only one controller influencing the central hub. Suppose the system starts in an initial state of $x_0=x_1=0$ perfectly aligned with influencer B. We thus see that in comparison to the scenario in which only A influences the star the presence of B's influence tends to extend the time scale at which A's influence becomes effective, i.e., we conclude that, in a system initially aligned with B's influence, B's best strategy to maximize its short term success is to shadow A. 

\subsection{Mean-field analysis of networks with bimodal degree distribution}
\label{MF} 
\begin{figure*}[tbp]
 \begin{center}
\includegraphics[width=.95\textwidth]{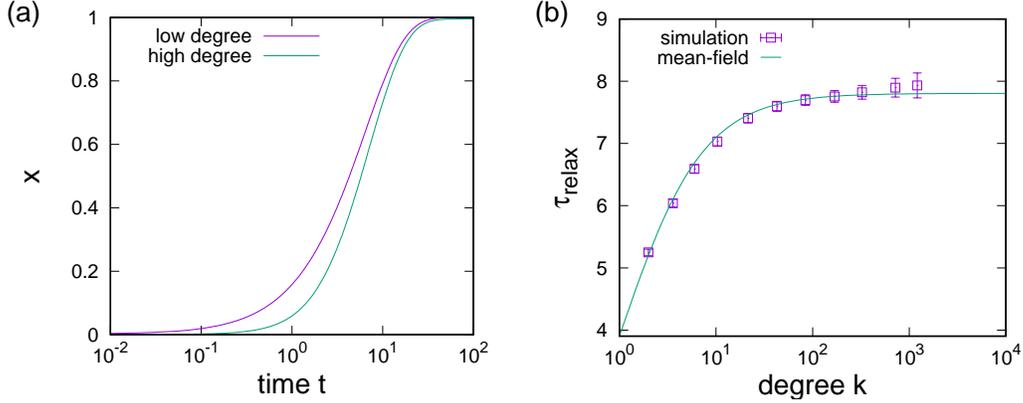}
 \caption {(a) Evolution of average opinions of low-degree and high-degree nodes over time. (b) Simulation estimates for the dependence of relaxation times of the opinion dynamics of nodes on their degree and comparison to the mean-field estimate of Eq. (\ref{ENN0}). Simulations were run on a network of $N=10^4$ nodes with power law degree distribution with exponent $\alpha=3$ and average degree $\langle k\rangle=5.8$ generated from a configuration model. For (a) trajectories are averaged over 200 independent runs and nodes with degree larger than 50 were labeled as high degree, whereas other nodes were labeled low-degree nodes. Initial conditions were all opinions opposed to A and influence of strength one was applied to all nodes from A without any influence from B.}
\label{fig.0a}
\end{center}
 \end{figure*} 
 
In subsection \ref{Star}, for the relatively simple case of star networks, we have seen how time scales at which a node equilibrates depend on its degree. As a result of the slow equilibration dynamics of hub nodes, it thus became apparent that hub control will result in poor vote share gains in the short run. To add further support to this finding, we also present an additional argument based on a degree-based mean-field theory. To explore the effects of hub and periphery control, we then apply this mean-field argument to  a class of random networks with bimodal degree distribution.

To proceed, we group nodes by degree and rewrite Eq. (\ref{E1}) for the dynamics of nodes with degree $k$
\begin{equation}
\label{EN0}
\dot{x}_k=-x_k + 1/\Delta_k \langle \sum_j a_{ji}\rangle_k x_j + p_{A,k}/\Delta_k,
\end{equation}
where $p_{A,k}$ and $p_{B,k}$ now denote the amount of influence controller A or B apply to a node of degree $k$, $\Delta_k=k+p_{A,k}+p_{B,k}$ and $\langle \cdot \cdot \cdot \rangle_k$ stands for the average over all nodes of degree $k$. We then approximate $\langle \sum_j a_{ji} \rangle_k \approx k \langle x\rangle$, where $\langle x\rangle = \sum_k kp_k/\langle k\rangle x_k$ represents the "average field" experienced by a node and $p_k$ denotes the fraction of nodes with degree $k$. Multiplying Eq. (\ref{EN0}) by $kp_k/\langle k\rangle$ and summing over $k$, we obtain an equation for the dynamics of $\langle x\rangle$
\begin{equation}
\label{EN1}
 \frac{d}{dt} \langle x\rangle = \left( \sum_k k^2p_k/(\langle k\rangle \Delta_k)-1 \right) \langle x\rangle + \sum_k kp_k p_{A,k}/(\langle k\rangle \Delta_k).
\end{equation}
Equation (\ref{EN1}) gives a linear first order differential equation for $\langle x\rangle$ and can be solved by standard methods. Assuming, e.g., voters that are initially opposed to the controlling influence, we obtain $\langle x\rangle (t)=B(\{ p_{A,k}\})/A(\{p_{A,k}\})(1-\exp(-A(\{p_{A,k}\})t))$, where $A(\{p_{A,k}\}=1-\sum_k k^2p_k/(\langle k\rangle \Delta_k)$ and $B(\{ p_{A,k}\})=\sum_k kp_k p_{A,k}/(\langle k\rangle \Delta_k)$ and both $A$ and $B$ depend on network structure and the influence configurations of controllers A and B, but we have only emphasized the dependence on $p_{A,k}$ which is of importance for our later argument. Using the solution of Eq. (\ref{EN1}) we can find mean-field solutions for the relaxation dynamics of nodes of degree $k$ by solving Eq. (\ref{EN0}). To illustrate the effect of different relaxation times of nodes depending on degree, let us consider a scenario with $p_{B,k}=0$ and $p_{A,k}=1$ for all $k$. Again assuming $x_k(0)=0$ and using standard methods to solve this inhomogeneous linear differential equation, we find
\begin{equation}
x_k(t)=\frac{1}{1+1/k} \left( C_0 \exp(-t) + C_1 \exp{(-A t)} \right) + C_2, 
\end{equation}
with constants $C_0, C_1$ and $C_2$ which are functions of $A$ and $B$. Following \cite{Hong:2002,Brede:2010a}, we can approximate an overall relaxation time via $\tau_\text{relax}= \int_{t=0}^{\infty} \frac{x_k(t)-x_k(\infty)} {x_k(0)-x_k(\infty)} dt$ and obtain
\begin{equation}
\label{ENN0}
 \tau_\text{relax} \propto \frac{1}{1+1/k}
\end{equation}
as a mean-field estimate for the time scale of the relaxation dynamics of a node with degree $k$ in a degree-heterogeneous network. Thus, as for the star networks considered above, we again emphasize that relaxation times of nodes are increasing functions of their degrees. This point is also emphasized by numerical simulations illustrated in Fig. \ref{fig.0a}, where we explored the relaxation dynamics of scale-free networks on which every node is influenced with fixed strength by controller A whereas controller B does not exert any influence. In panel (a) we compare averaged trajectories of opinions of low-degree (degree $\leq 50$) and high degree (degree $>50$) nodes. As expected the numerics reveal that the relaxation of hub nodes is markedly slower than the dynamics of low-degree nodes. More in detail, in panel (b) we estimated the dependence of relaxation times on degree. Results are in very good agreement with the mean field estimate of Eq. (\ref{ENN0}).

So far in this section, we have shown that the relaxation of hub nodes is typically slower than that of low-degree nodes, but have not specifically related these results to scenarios of hub or periphery control, which is the aim of the remainder of this section. Equation (\ref{EN1}) gives the dynamics of the mean field encountered at the end of a randomly chosen link, but we can derive a similar relationship for overall vote shares $X_A=1/N \sum_{i=1}^N x_i=\sum_k p_k x_k$ by multiplying Eq. (\ref{EN0}) by $p_k$ and summing over $k$ to obtain
\begin{equation}
\label{EN2}
 \frac{d}{dt} X_A =-X_A +\langle x \rangle (t) \sum_k kp_k/\Delta_k + \sum_k p_k p_{A,k}/\Delta_k,
\end{equation}
which gives another linear first order differential equation for the evolution of $X$. For initial condition $X(0)=0$ equation (\ref{EN2}) is solved by
\begin{align}
\nonumber
 X_A(t)=&\frac{C (\{p_{A,k}\}) B(\{p_{A,k}\})}{A(\{p_{A,k}\})} + D (\{p_{A,k}\})+ \left( \frac{C(\{p_{A,k}\})B(\{p_{A,k}\})}{1-A(\{p_{A,k}\})}-D(\{p_{A,k}\}) \right) e^{-t} -\\ \label{EN3}
 &-\frac{C(\{p_{A,k}\})B(\{p_{A,k}\})}{A(\{p_{A,k}\})(1-A(\{p_{A,k}\}))} e^{-A(\{p_{A,k}\})t},
\end{align}
where $C(\{p_{A,k}\})=\sum_k kp_k/\Delta_k$ and $D(\{p_{A,k}\})=\sum_k p_k p_{A,k}/\Delta_k$. One notes that generally $1-1/(1+\frac{p_A+p_B}{\langle k\rangle})) \geq A(\{p_{A,k}\}) \geq 0$ (see Appendix ?), where $p_A=\sum_k p_k p_{A,k}$ and $p_B=\sum_k p_k p_{B,k}$ are the average control allocations per node of controllers A and B. Thus, the slowest time scale in the relaxation dynamics of $\langle x\rangle(t)$ is given by $e^{-At}$. Since also $B(\{p_{A,k}\}) >0$ and $C(\{p_{A,k}\})>0$, the last term in Eq. (\ref{EN3}) reflects a negative contribution on vote shares which is gradually compensated for as the influence of controller A spreads over the network. In contrast, the first two summands in Eq. (\ref{EN3}) relate to equilibrium vote shares, and the term proportional to $e^{-t}$ describes a fast relaxation dynamics whose time scales are not influenced by control allocations or network structure.

For a better understanding of effects of control on short time scales we focus on the last term of (\ref{EN3}) in the following and highlight differences between hub and periphery control. Detailed results will generally depend on the details of control allocation of A and B. However, to illustrate differences between hub and periphery control we consider a random network composed of equal numbers of nodes of degrees $k_1$ and $k_2>k_1$ and compare the effects of allocating control of strength $p>0$ exclusively to either all of the hub nodes with degree $k_2$ or to all of the periphery nodes of degree $k_1$. Let us further assume that all nodes are subject to opposing influence of strength $q>0$ from controller B. We thus have
\begin{align}
A_H&=1-\frac{1}{k_1+k_2} \left( \frac{k_1^2}{k_1+q} + \frac{k_2^2}{k_2+p+q} \right) \\
B_H&=pk_2 \frac{1}{k_1+k_2} \frac{1}{k_2+p+q}\\
C_H&=1/2 \frac{k_1}{k_1+q}+1/2 \frac{k_2}{k_2+p+q}
\end{align}
for a scenario of exclusive hub control and
\begin{align}
A_P&=1-\frac{1}{k_1+k_2} \left( \frac{k_1^2}{k_1+p+q} + \frac{k_2^2}{k_2+q} \right) \\
B_P&=pk_1 \frac{1}{k_1+k_2} \frac{1}{k_1+p+q}\\
C_P&=1/2 \frac{k_1}{k_1+p+q}+1/2 \frac{k_2}{k_2+q}
\end{align}
for exclusive periphery control. We next note that $A_H>A_P>0$, i.e. the slowest relaxation dynamics is faster when focusing control on hub nodes. However, also $C_H B_H/(A_H (1-A_H)) \geq C_P B_P/(A_P (1-A_P))$ irrespective of $p$ and $q$ as long as $k_2>k_1>0$, i.e. control allocation to hub nodes generally incurs a larger initial loss of vote shares than control allocation to periphery nodes. We thus see that vote shares are larger for periphery control at short time scales and larger for hub control at longer time scales as the negative term decays faster in that case.

\subsection{Analysis of scale-free networks}
\label{sec:2}
\begin{figure*}[ht]
 \begin{center}
\includegraphics[width=.95\textwidth]{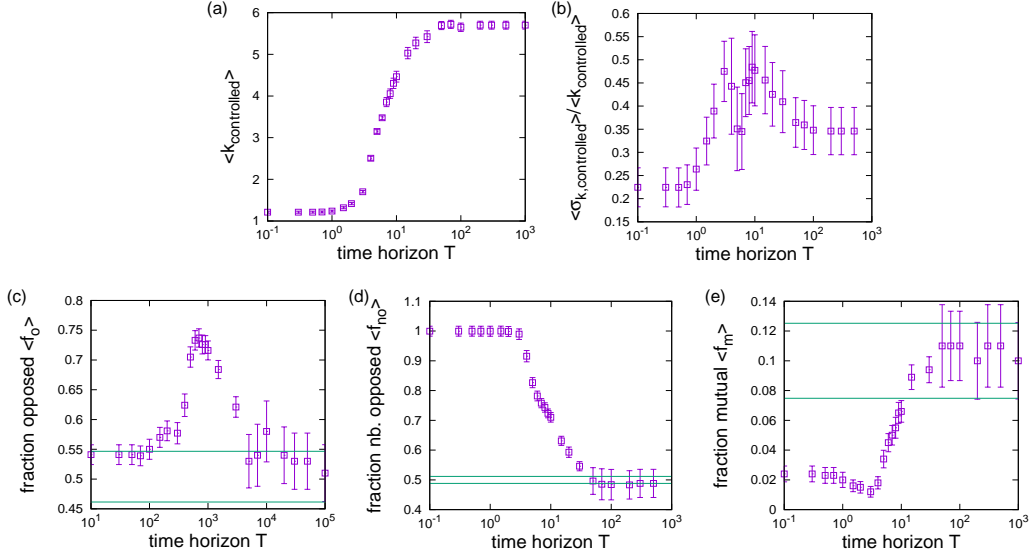} 
 \caption {Dependence of optimal influence strategies on time horizons T of the influence maximizer: (a) averaged influenced degrees, and (b) relative standard deviation of distribution of influenced degrees. Bottom row: (c) fraction of opposing initial opinions, (d) fraction of neighbors of influenced nodes which have opposing initial conditions, and (e) fraction of mutually influenced votes. Settings are $n_A=n_B=10$ and $q=0.5$, i.e., $50\%$ of initial votes are A or B. The data are estimated from 100 runs for $t<30$ and 10 runs otherwise. The lines in (b) and (c) indicate a range of one standard deviation around the expectation for a random allocation.}
\label{fig.2}
\end{center}
\end{figure*}

In the following we present numerical experiments conducted for networks with scale-free degree distributions. We start by illustrating the effectiveness of optimization, and plot the dependence of the gain achievable by optimization on the amount of resource available to the active voter for various time horizons of the optimizer (see Fig. \ref{fig.0}a). It becomes clear that optimization gains strongly depend on both the optimizer's time horizon and its resource availability, and we can make two observations. First, the shorter the time horizon, the less exploitable is a passive strategy, and maximum optimization gains are limited to at most $5\%$ for the shortest time frame we investigated ($T=1$). Second, strategic allocation is the more important the less resource available to the influencer, so that gains are largest at around 50\% for an influencer who can influence $n_A=1$ node against $n_B=10$. 

In panel Fig. \ref{fig.0}b we present data for the optimized average vote share trajectories $X_A(t)$ for $n_A=n_B=10$, for situations optimized for short and long time horizons $T=1,5,30$ and $T=1000$. The figure illustrates that it is indeed possible to achieve short term vote share increases, but they come at a cost of long term reductions in vote shares (e.g., for $T=1$ or $T=5$). In contrast, far more substantial long-term vote share gains (e.g, for $T=30$ or $T=1000$) are also possible, but such gains come at the cost of short term vote share losses (observe the dips at around $t=50$ for $T=30$ and $T=1000$). 

In panel Fig. \ref{fig.0}c we also plot the fraction of influenced nodes as a function of degree for short, intermediate, and long term control time horizons. It becomes clear that strategic control strategies differ substantially: whereas optimal control strongly focuses on leaf nodes for $T=1$, nodes of intermediate degree are targeted for $T=5$ and control again focuses almost exclusively on hub nodes for long time horizons as for $T=30$ as already observed in other work \citep{Masuda:2015}. Together with our results of long transients to control hub nodes in Sec. \ref{Star} we can thus understand the initial decline in vote shares for long term control: the passive influencer who targets a random selection of nodes can gain short term vote shares while influence gains of the long term optimizer are initially limited when control of hub nodes has not yet been achieved.

We proceed with a more systematic investigation of the dependence of optimal influence strategies on the strategic influencer's time horizon. For this purpose we have run experiments for varying $T$ and have recorded statistics about nodes targeted by the optimized control for each optimized configuration. As before, a key quantity of interest is the degree of controlled nodes, and, consequently, we measure $\langle k_\text{controlled}\rangle=\sum_i p_{A,i} k_i/n_A$ where $k_i$ denotes the degree of node $i$. To gain insights about the distribution of influenced degrees we also record its standard deviation $\langle \sigma_\text{k,controlled}^2\rangle=1/n_A \sum_i p_{A,i} (k_i-\langle k_\text{controlled}\rangle)^2$. Further, we are interested to measure how strategically optimized control relates to nodes influenced by the passive influencer and thus measure the fraction $\langle f_m\rangle=1/n_A \sum_i p_{A,i} p_{B,i}$ of nodes influenced by A which also experience control by B. For randomly allocated influence B one expects $\langle f_m\rangle_\text{rand}=n_B/N$. Thus, $\langle f_m\rangle>\langle f_m\rangle_\text{rand}=n_B/N$ indicates crowding of A influence on B-influenced nodes whereas $\langle f_m\rangle>\langle f_m\rangle_\text{rand}=n_B/N$ characterizes avoidance of B controlled nodes by A. 

Since the stationarity condition of Eq. (\ref{E1}) is a linear equation it allows for only one solution  and hence strategies for optimal stationary influence maximization are independent of initial vote allocations. Importantly, this does not need to be the case for optimal allocations subject to limited time horizons. We thus test for correlations of control allocations with initial vote configurations $\{x_i(0)\}_{i=1}^N$ by measuring the average number of influenced nodes with initially opposed votes $\langle f_o\rangle=1/n_A \sum_i p_{A,i} \delta_{B,x_i(0)}$ where $\delta_{x,y}=1$ if $x=y$ and $\delta_{x,y}=0$ otherwise (Kronecker delta). Hence,  $\langle f_o\rangle<1-q$ indicates avoidance of controlling initially opposed votes, whereas for $\langle f_o\rangle>1-q$ optimal control focuses on influencing initially opposed votes. Similarly, we are also interested in local neighborhoods of controlled nodes. Consequently, we measure the fractions of initially opposed votes amongst an influenced nodes' neighbors by defining an opposed-neighbor ratio $\langle f_{no}\rangle=1/n_A \sum_i p_{A,i} 1/k_i \sum_j a_{ji} \delta_{x_j(0),B}$. As above, $\langle f_{no}\rangle<1-q$ indicates avoidance of influencing nodes with many opposed neighbors and $\langle f_{no}\rangle>1-q$ suggests that  control is focused on controlling nodes with opposed neighbors.

\begin{figure*}[tbp]
 \begin{center}
\includegraphics[width=.95\textwidth]{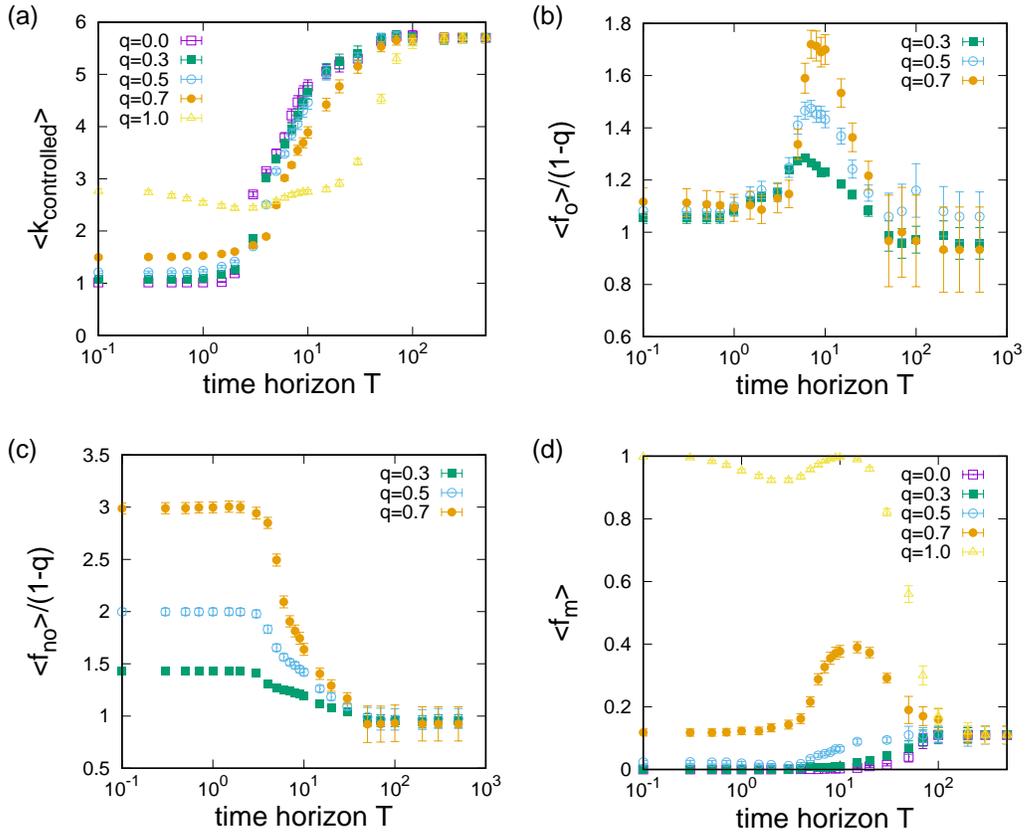}
 \caption {Dependence of optimal influence strategies on time horizons T of the influence maximizer for different initial configurations characterized by the initial fraction of aligned votes q: (a) averaged influenced degrees, (b) normalized fraction of opposing initial opinions, (c) normalized fraction of opposing neighbors of influenced nodes, and (d) fraction of mutually influenced votes. Settings are $n_A=n_B=10$, $\alpha=3$, $N=100$. The data are estimated from 100 runs for $t<30$ and 10 runs otherwise. Note, that there are no curves for $q=0$ and $q=1$ in (b) and (c), as correlations with initial conditions are meaningless for homogeneous initial conditions.}
\label{fig.3}
\end{center}
\end{figure*}

Numerical result representing averages over 100 network and initial vote allocation configurations are illustrated in Fig. \ref{fig.2}. The top row of panels in Fig. \ref{fig.2} plots characteristics of the distribution of influenced degrees vs. optimization time horizons. From panel Fig. \ref{fig.2}a we see that there is indeed a  transition between a regime of lowest-degree node control and a regime of hub control, leaving only a limited range of time horizons for which intermediate degree control is optimal.  Plots of the relative standard deviations in Fig. \ref{fig.2}b show that control tends to be most focused on low (high) degree nodes for short (long) time horizons, with least focus for intermediate time scales when both high and low degree nodes are targeted in the optimal control allocation.

In the bottom row of panels of Fig. \ref{fig.2} we present further results on changes of correlations of optimal control allocations with initial conditions and the opponent's influence allocation with time horizons. As expected, $\langle f_o\rangle \approx q=0.5$ and $\langle f_{no}\rangle \approx q=0.5$ for $T \gg 1$, i.e., no statistically significant correlations with initial configurations are observed when time horizons approach the stationary scenario. In contrast, for small $T$ and for intermediate $T$ respectively, marginal and very strong correlations are found.  More specifically, we see that optimal allocations mostly target nodes with opposing neighbors for very small $T$ (Fig. \ref{fig.2}d), an effect that gradually diminishes as $T$ is increased. For intermediate time scales, the strategic influencer no longer focuses on opposed neighbors of controlled nodes, but instead selects controlled nodes based on their own initial states (see Fig. \ref{fig.2}c). Thus, whilst for small $T$ it mostly matters to influence a low degree node with initially opposed neighbors for intermediate $T$ the focus should be on influencing nodes whose initial votes differ from the influencers' target. Finally, panel Fig. \ref{fig.2}e shows data for the variation of correlations with the opposing controllers' influence $\langle f_m\rangle$ with $T$. We can again distinguish three regimes. For small $T$ we have $\langle f_m\rangle<0.1=n_B/N$ and thus strategic control is found to avoid nodes influenced by the passive influencer. This effect diminishes for intermediate $T$ and finally for close to stationary control no significant correlation between targeted nodes and passively controlled nodes is found.

These results suggest that optimal influence allocations for short and intermediate time horizons are strongly influenced by the initial configuration of votes on the social network. To explore this effect further we carried out numerical experiments in which we varied the fraction of initially aligned votes $q$. Results are summarized in the panels of Fig. \ref{fig.3}. As above in Fig. \ref{fig.2}, we analyze the average controlled degree $\langle k_\text{controlled}\rangle$, the fraction of influenced voters with initially opposed votes and fraction of initially opposed neighbors of influenced nodes (which we normalize now by $1-q$ to account for the varying fraction of opposed votes), and the fraction of mutually influenced votes. We observe that all quantities show a significant dependence on $q$ as long as time horizons are short, but, as expected, $q$-dependencies vanish when approaching stationary control.

In more detail, Fig. \ref{fig.3}a shows that the transition between optimal low and high degree control systematically depends on $q$. The larger the fraction of initially aligned votes, the larger the low degree control regime, but also the less significant the difference between optimally targeted nodes in both regimes (cf. the curve for $q=1$ in Fig. \ref{fig.3}a). In fact, as for $q=1$ the control allocation almost exclusively focuses on shadowing the opposing controller, no focused lowest-degree control regime exists. Further, in Fig. \ref{fig.3}b the importance of targeting opposing votes is shown to also be a function of time horizons and initial conditions. In particular, we notice that, for intermediate time horizons, controlling initially opposed votes becomes the more important the less initially opposed votes are present in the initial state. Again, controlling nodes with initially opposed neighbors is only relevant for very short time horizons when only leaf nodes are targeted and the transition from this regime towards a regime in which opposing-neighbor control becomes irrelevant is roughly independent of initial conditions (see Fig. \ref{fig.3}c and notice that for $T<5$ in all cases only leaf nodes with an initially opposing neighbor are targeted).

Finally, in Fig. \ref{fig.3}d we investigate the relevance of neutralizing the passive influencer by shadowing its allocation as measured by $\langle f_m\rangle$. Depending on time horizons and initial state, three regimes exist. First, for large time horizons we observe $\langle f_m\rangle=0.1=n_B/N$, i.e. optimal allocations are neutral relative to the passive control. However, for short and intermediate time horizons this is different. In a second regime, for short and intermediate $T$ and relatively low numbers of initially unaligned votes (roughly $q<0.5$) we find $\langle f_m\rangle \approx 0$. In other words: in this regime optimal control avoids the passive controller. In contrast, if a larger fraction of roughly $q \geq 0.7$ initial votes is already aligned with the active controller $\langle f_m\rangle>0.1=n_B/N$, i.e. optimal strategies aim to target the passive control. This is particularly the case in at intermediate time horizons (cf. curves for $q=0.7$ and $q=1.0$). In the extreme case of initial conditions perfectly aligned with the active controller for $q=1$ shadowing becomes particularly evident: here the best strategy clearly is to neutralize the passive controller by perfectly shadowing nodes influenced by her. As our model does not allow for perfect neutralization, the influence of the passive controller will always penetrate the system on very long time scales and thus perfect shadowing of the passive controller will not necessarily make sense for stationary control. However, aiming at shadowing the passive controller for short time horizons can extend the time scale at which its influence can propagate and is thus a viable strategy (see also Sec. \ref{Star}). 

\begin{figure*}[tbp]
 \begin{center}
\includegraphics[width=.95\textwidth]{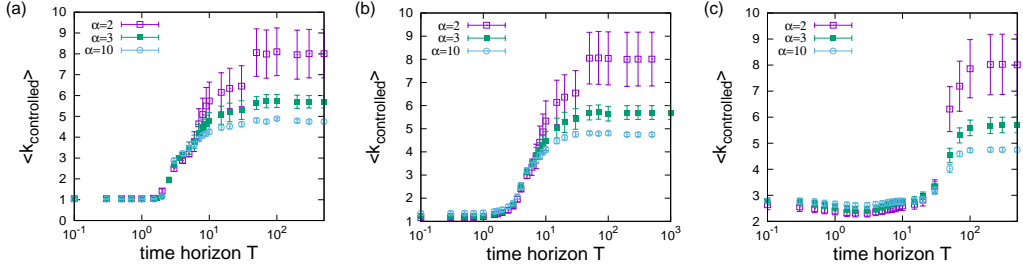}
 \caption {Dependence of average controlled degree $\langle k_\text{controlled} \rangle$ on time horizons of the influencer for different degrees of network heterogeneity, i.e. $\alpha=2,3$ and $\alpha=10$ (but $\langle k\rangle=3$ for all $\alpha$). Panels (a) to (c) present data for initial vote share allocations with $q=0$ (a), $q=0.5$ (b), and $q=1.0$ (c). Other settings are $n_A=n_B=10$, $N=100$. The data are estimated from 100 runs for $t<30$ and 10 runs otherwise.}
\label{fig.4}
\end{center}
\end{figure*}

Hence, we see that the reasons for low degree control are twofold, but both are related to time scales at which influence over certain nodes on the social network can be gained. As illustrated in our example of a star network in Sec. \ref{Star}, one reason relates to the long transient time required to gain influence over high degree nodes. As just demonstrated, the second reason relates to extending the time scales at which an opposing influencer can gain control over a node already aligned with the active influencer which can be extended by opposing the other influencers control. The relative relevance of both aspects depends on initial conditions: when many initial votes are not aligned, the first aspect becomes the dominant consideration. However, if many nodes are initially aligned with the controller, then the second consideration becomes paramount.

As a last point in this section we address the dependence of the optimal low- and high degree control schemes on network heterogeneity. For this purpose we construct networks with varying degree exponent $\alpha$ as described in Sec. \ref{Model}. Results for different initial vote allocations are given in the panels of Fig. \ref{fig.4}. Cross comparison of panels (a)-(c) which represent different $q$ show that transitions are dependent on initial conditions, generally allowing for an extended low-degree control regime when initial votes are mostly aligned with the strategic controller. In contrast, apart from an expected rise in the average maximum degree when varying $\alpha$, comparison of very heterogeneous networks with $\alpha=2$ and almost regular networks for $\alpha=10$ does not show a significant difference for all investigated $q$. Low- and high-degree control regimes are thus largely independent of network heterogeneity.

\subsection{Heuristics}
\label{sec:3}

\begin{figure*}[tbp]
 \begin{center}
\includegraphics[width=.95\textwidth]{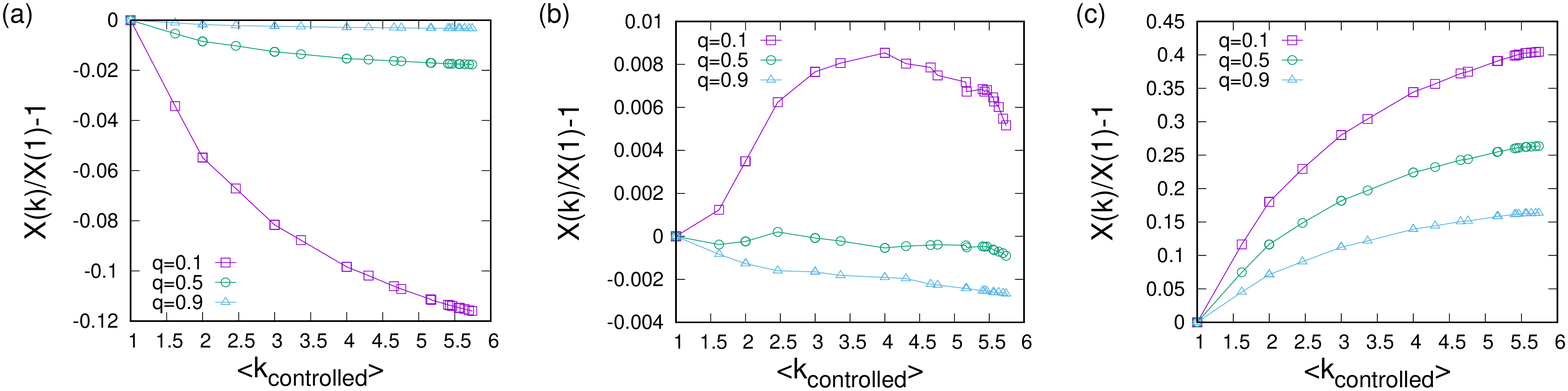} 
 \caption {Dependence of vote shares at (a) very short ($T=1$), (b) intermediate ($T=4.5$) and (c) long ($T=50$) time scales on average degrees of controlled nodes on scale-free networks with $\alpha=3$, $\langle k\rangle=3$ of size $N=10^4$. Other settings are $n_A=n_B=10^3$ and data points represent averages over 10 runs.}
\label{fig.3a}
\end{center}
\end{figure*}

Above, we have investigated optimal control strategies by numerical optimization of votes shares for relatively small networks and the question arises if these results for small systems generalize to larger networks for which optimization is not computationally feasible with limited resources. To address this issue we have evaluated heuristics for strategies on larger networks. For this purpose, we assign nodes scores based on their degree, initial conditions, correlations with initial conditions of neighbors, and correlations with the passive controller 
\begin{equation}
\label{EqS}
S_i=\mid k_i-k_\text{target}\mid - a_1 x_i^\text{nb}(0) - a_2 \delta_{x_i(0),B}  - a_3 \delta_{p_{A,i},p_{B,i}}
\end{equation}
where $x_i^\text{nb}(0)=1/k_i \sum_j a_{ji} \delta_{x_j(0),B}$ is the fraction of B-neighbors of $i$ and $k_\text{target}$ gives the degree of nodes to be targeted.  The coefficients $a_j$, $j=1,2,3$ allow to tune the relative influence of various criteria in the heuristic and we typically set $a_j=1$ (or $a_j=-1$) if criterion $i$ is accounted for as a positive or negative influence in the heuristic (or $a_j=0$ if the criterion is ignored).  The choice of $a_j \in \{-1,0,1\}$ allows for limited trade-off in degree heterogeneity to allow meeting other criteria and results presented below are robust for other choices of small values of $a_j$ and we have not observed any situation in which choices of large values of $a_j$ would result in markedly superior performance.

\begin{table}
\label{tab.1}
\begin{center}
\begin{tabular}{c c c c c}
\hline
\multirow{2}{*}{Heuristic ($a_1,a_2,a_3$)} & \multicolumn{2}{c}{$T=1.0$} & %
    \multicolumn{2}{c}{$T=4.5$} \\
\cline{2-5}
 & $q=0.1$ & $q=0.9$& $q=0.1$ & $q=0.9$ \\
\hline
$(1,0,0)$  &$0.16\pm 0.01 $ & $2.3\pm 0.1$   & $0.8\pm 0.1$ & $4.8\pm 0.2$\\
$(-1,0,0)$ &$-0.16\pm 0.01$ & $-0.9\pm 0.1$  & $0.2\pm 0.1$ & $-1.1\pm 0.1$\\
$(0,1,0)$  &$0.03\pm 0.01 $ & $2.1\pm 0.1$   & $0.4\pm 0.1$ & $1.9\pm 0.2$\\
$(0,-1,0)$  &$-0.10\pm 0.01 $ & $0.0\pm 0.1$ & $0.0\pm 0.1$ & $0.4 \pm 0.2$\\
$(0,0,1)$  &$-0.16\pm 0.01 $ & $0.9\pm 0.1$  & $-0.1\pm 0.1$& $3.7 \pm 0.2$\\
$(0,0,-1)$  &$0.06\pm 0.01 $ & $-0.3\pm 0.1$ & $0.5\pm 0.1$ & $-1.1 \pm 0.2$\\
\hline
\multirow{2}{*}{best} &$(1,1,-1)$    &$(1,1,1)$    & (1,1,-1)     & (1,1,1)\\
                      &$0.24\pm .02$ &$3.1\pm 0.1$ & $1.2\pm 0.1$ & $8.1\pm 0.3$\\
\hline
\end{tabular}
\end{center}
\caption{Evaluation of heuristics based on Eq. (\ref{EqS}) for different initial fractions of aligned votes $q$ and very short and intermediate time horizons $T=1.0$ and $T=4.5$. Values in the table give the maximum achievable gain of the heuristic (when varying $k_\text{target}$) relative to the range of variation observed in purely degree-based heuristics, i.e. $0.16$ for heuristic $(1,0,0)$ for $q=0.1$ means that the heuristic can achieve an improvement of 16\% relative to improvements that can be achieved by tuning average targeted degrees alone. The last row lists best heuristics and gives relative gains achievable. The data are based on simulations performed for networks of size $N=10^4$ for which experiments with varying $k_\text{target}$ have been run, cf. Fig. \ref{fig.3a} for each heuristic.}
\end{table}

For instance, we have evaluated  purely degree-based heuristics (i.e. $a_1=a_2=a_3=0$) by systematically varying $k_\text{target}$ and evaluating vote shares at different time horizons. Results for different initial conditions and short, intermediate, and long time horizons are summarized in Fig. \ref{fig.3a}, in which we plot the dependence of normalized vote share differences $X(k)/X(1)-1$ on the average controlled degree which we have tuned by varying $k_\text{target}$, where $X(1)$ denotes vote shares when only nodes of degree one are influenced. Confirming our results above, it becomes clear that a strategy of lowest-degree control is always superior for short time horizons (cf. Fig. \ref{fig.3a}a) whereas highest-degree control is always optimal in the long run (cf. Fig. \ref{fig.3a}c). For both time scales relative gains for the best allocations can be substantial with the largest differences occurring when the influence of the active controller in the initial state is small. In contrast, for intermediate time scales relative differences of degree-based heuristics are very small. Nevertheless, we also observe that, for intermediate time scales, optimal control can be achieved by targeting nodes of intermediate degree (cf. Fig. \ref{fig.3a}b for $q=0.1$. Similar peaks can be observed for other values of $q$ but at different $T$).

For a more comprehensive evaluation of the proposed heuristics, we have run simulations to evaluate the dependence on $k_\text{target}$ for all 27 heuristics for different initial conditions and short and intermediate time horizons. We have not investigated very long time horizons, as no significant correlations are expected in this case. Results of gains relative to gains achievable by target-degree tuning are summarized in Tab. 1, where we list results for each individual factor in Eq. (\ref{EqS}) and also give the best combination of factors for each parameter setting. In agreement with observations above, we always find a positive influence of $a_1$, i.e., targeting nodes surrounded by many opposed votes enhances vote share gains for short and intermediate time scales, whereas seeking out nodes surrounded by many aligned votes has the opposite effect. A smaller positive effect is found for $a_2$ parameterizing correlations with initial conditions of the targeted node itself. Effects of opponent shadowing measured by $a_3$, on the other hand,  depend on initial conditions. Confirming our earlier results in subsection \ref{sec:2} we find positive effects of opponent avoidance for non-aligned initial settings ($q=0.1$), and positive effects of opponent shadowing for largely aligned initial settings ($q=0.9$).

\subsection{Dynamic heuristics}
\label{sec:4}
\begin{figure*}[tbp]
 \begin{center}
\includegraphics[width=.95\textwidth]{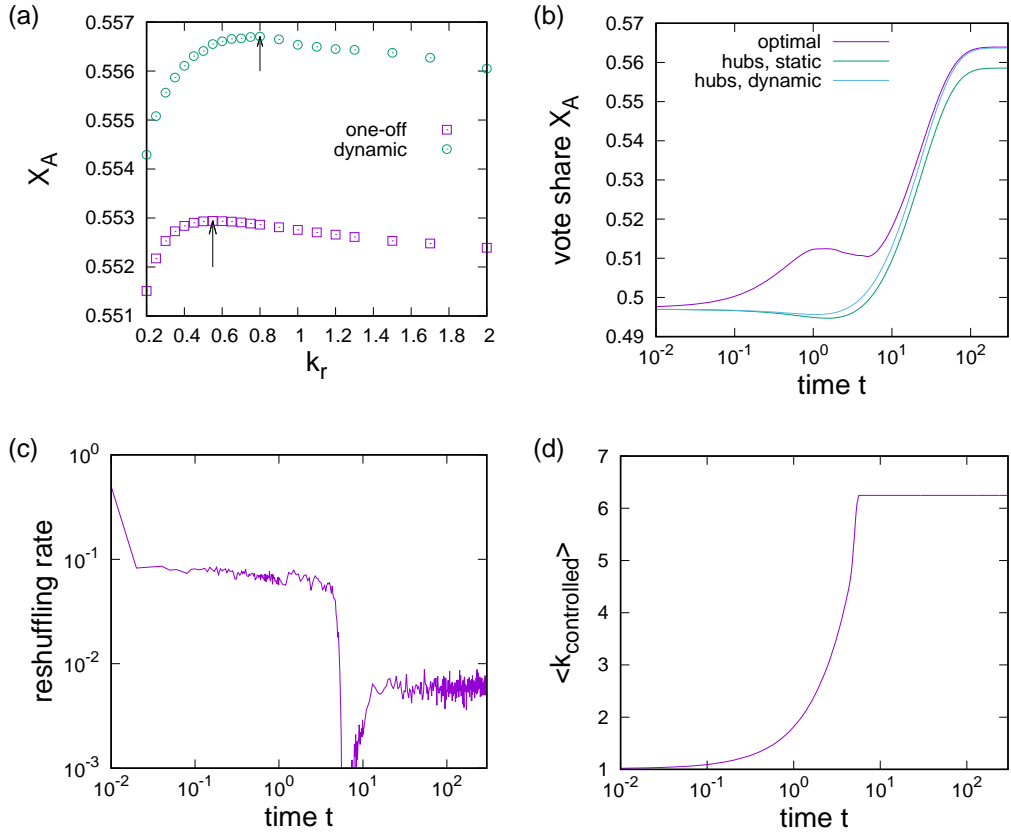} 
 \caption {(a) Evaluation of dynamic heuristics: dependence of time averaged vote shares on growth rate of average degree $k_r$ for a dynamic heuristic based on initial settings ("one-off") and a heuristic based on reallocations making use of projections from the master equation ("dynamic"). The arrows indicate the maxima for $k_r=0.55$ and $k_r=0.8$. For comparison, note that for one-off hub allocation one finds $X_A=0.550\pm 0.001$ and for dynamic hub allocation one has $X_A=0.555\pm 0.001$. (b) Comparison of the evolution of vote shares for a static allocation to hubs, allocation to hub nodes subject to dynamical re-allocation, and using the optimal dynamical degree heuristic. (c,d): Dependence of the average number of vote-share based reallocations and average controlled degree on time for the optimal dynamic degree-dependent heuristic. For all experiments show initial conditions are balanced ($q=0.5$) and trajectories represent averages over 100 runs for a system of size $N=10^3$ with $n_A=n_B=10^2$ for scale-free networks with $\alpha=3$ and $\langle k\rangle=6$.}
\label{fig.5}
\end{center}
\end{figure*}

Our main finding suggests that, when deciding about a one-off selection of nodes to target, influence-maximizing strategies should select low-degree nodes at short time horizons and high degree nodes for long time horizons. The natural question arises whether a dynamic allocation scheme that re-assigns targeted nodes over time could combine the short term benefits of low-degree allocations with the long-term gains of hub allocations?  

To address this question, we will consider dynamic influence allocation schemes in this section. We will consider two degree-based settings. In a first setting, initially the $n_A$ lowest degree nodes are targeted and then control is systematically shifted towards larger degree nodes as the dynamics of voting unfolds. More in detail, as we numerically integrate Eq. (\ref{E1}), in each integration step with probability $k_r$ control is removed from the controlled node with the lowest degree and rewired to the uncontrolled node with the smallest degree that is higher than the degree of the previously controlled node. In this way average degrees of controlled nodes are increased at a rate proportional to $k_r$ over time. Second, we consider a targeting scheme in which the $n_A$ highest degree nodes are targeted. 

Both of the above two schemes generally allow for some degeneracy as multiple different nodes fulfill these purely degree-based criteria. We further select the nodes to be targeted among all nodes of degree equal to the selected nodes, based on fractions of uncontrolled neighbors either in the initial or estimated current state of the system based on Eq. (\ref{E1}). We thus compare the following static and dynamic schemes:
\begin{itemize}
\item (static hubs): targeting hubs, resolving degeneracy by selecting nodes with largest fractions of initially opposed neighbors,
\item (dynamic hubs): target hubs, resolve degeneracy by selecting nodes with largest estimated average deviation of neighbors $1/k_i \sum a_{ji} x_j(t)$ from the controller,
\item (increase degree, one-off): increase average degree at rate $k_r$, resolve degeneracy by targeting nodes with largest fraction of initially opposed neighbors,
\item (increase degree, dynamic): increase average degree at rate $k_r$, resolve degeneracy by selecting nodes with largest estimated average deviation of neighbors $1/k_i \sum a_{ji} x_j(t)$ from the controller.
\end{itemize}
The increasing-degree schemes require tuning of the rate of increase of degrees $k_r$. Clearly, if $k_r$ is too large, the scheme quickly shifts to hub control and effects of low-degree control cannot become effective. On the other hand, if $k_r$ is too small, longer term losses of vote shares as observed in Fig. \ref{fig.0}b might ensue. The resulting trade-off is analyzed in Fig. \ref{fig.5}a in which we plot simulation results for the dependence of time averaged vote shares
\begin{equation}
 \overline{X}_A= 1/(NT)\int_0^T \sum_i x_i(t) dt
\end{equation}
vs. the rate of increase of degrees $k_r$ for a one-off and a dynamic increasing degree scheme. We typically chose $T=300$ which is long enough for system composed of $10^3$ nodes to reach a stationary state, but not too long as to hide effects from the transient dynamics (cf. Fig. \ref{fig.5}b). Even though differences in time-averaged vote shares turn out to be very small, a clear maximum can be identified for both cases and we notice that vote shares achievable in dynamic schemes can indeed be slightly larger than vote shares obtained when statically targeting hubs (maximum vote shares are $\overline{X}_A=0.553 \pm 0.001$ and $\overline{X}_A=0.557 \pm 0.001$ vs. $\overline{X}_A=0.550 \pm 0.001$). Further, we observe that degeneracy resolution based on correlations with estimated current states result in improved performance ($\overline{X}_A=0.557\pm 0.001$ vs. $\overline{X}_A=0.553 \pm 0.001$ for increasing degree schemes and $\overline{X}_A=0.555 \pm 0.001$ vs. $\overline{X}_A=0.550 \pm 0.001$ for dynamic or static hub targeting). 

In Fig. \ref{fig.5}b, we compare the resulting average trajectories of vote shares over time for the optimal degree-increasing dynamic scheme are compared with trajectories for hub-based static and dynamic schemes, and we notice two effects of interest. First, degree increasing schemes can indeed avoid the initial drop in vote shares and instead result in a slight initial gain (compare Fig. \ref{fig.5}b with Fig. \ref{fig.0}b). Second, by adaptively seeking out high-degree nodes with largest opposed neighbor fractions, dynamic targeting can also improve stationary performances. Panels (c) and (d) of Fig. \ref{fig.5} further show the change in the average controlled degree and average numbers of degeneracy resolution moves for the optimal dynamic degree-increasing scheme. We roughly see two different regimes in the plot of the average number of degeneracy resolution moves (Fig. \ref{fig.5}c): (i) whilst degrees are increasing control of roughly $10$ and (ii) when degrees have been allocated to the highest degree nodes control of roughly $1$ controlled nodes of equal degree are changed based on vote share estimates. We thus see that vote shares in the stationary state which are enhanced in comparison to static settings  indeed result from an ongoing process of adaptive control re-assignment.

\section{Discussion}

Previous literature on influence maximization in the voter model \citep{Yildiz:2013,Kuhlman:2013,Masuda:2015} has assumed long time horizons and thus  been restricted to the analysis of stationary states of the voting dynamics. However, in real-world settings influence maximizers may often have limited time horizons which might be very short in some situations. Accordingly, in this paper we have analyzed the effect of an influence maximizer's time horizon on optimal strategies for strategic influence maximization in the usual two-state voter model. 

Our analysis makes clear that influence maximization strategies depend on time horizons: Whereas long time horizons require mainly hub control, for short time horizons control is optimized by targeting low degree nodes. Using the example of a star network and developing a degree-based mean-field theory we have shown that the shift in control regimes is caused by a dependence on degree of effective time scales a controller needs to gain control over a node. Time scales to control hub nodes are long whereas control of low degree nodes can be reached more quickly.

As a second contribution, we have shown that in limited time-horizon control knowledge of the system's initial state can be exploited and optimal control strategies differ depending on details of the initial vote allocation and the exact time horizon of the optimizer. We have found three general rules of thumb to characterize optimal influencing strategies for the short and intermediate time horizon.  First, we have argued that when many initial votes are aligned with the controller, the controller should aim to neutralize opposing influence. Second, if many initial votes differ from the controller's target, a controller is best served by targeting low-degree nodes with opposing initial vote. Interestingly, similar to findings for synchrony-enhancing arrangements of anti-correlated native frequencies in diffusively coupled oscillator populations \cite{Brede:2008a,Brede:2008b,Brede:2010a}, such a strategy maximizes the corresponding interaction terms in the equation of motion. Third, for short time horizons controllers should focus on targeting nodes with large fractions of neighbors with opposed initial votes.  Having identified these heuristics optimizing control for small networks, we have also demonstrated their effectiveness for larger networks.

It is interesting to note that, in spite of very different modeling assumptions, our results agree with the findings of Alshamsi et al., who analyzed influence maximization for complex contagion \citep{Alshamsi:2017}. It may be arguable if the voter model can be considered a model of complex contagion as one single exposure is enough to change the state of a voter. However, on aggregate, a voter has higher likelihoods of assuming an opinion the more of its neighbors already hold that opinion, so that it might reflect the repeated exposure criterion in some averaged sense. Notwithstanding these differences, in both models, time scales to gain control of a node depend on the node's degrees and thus the balance of optimality changes from low-degree to hub control depending on how far influence has already penetrated the system. 

With our findings we have demonstrated that for one-off degree-based control allocations optimal long term control will necessarily suffer losses in vote shares in the short term. As we have shown, such losses can be avoided, when making use of dynamic allocations which target nodes with different degrees over time. More precisely, average targeted degrees of nodes should be increased at an intermediate rate which is low enough to allow gains from low-degree effects to become effective at short time scales, but fast enough to avoid losses from suboptimal targeting in the stationary state. Further, our results have shown that additional improvements can be made for the optimal degree-changing dynamics, by selecting between nodes of equal degree, and dynamically switching to nodes with largest average neighbor deviation in estimated vote shares to the controllers target state. 

Our study above has been limited to undirected networks, but some conclusions for the more general case of directed networks are straightforward. Our analysis in Sec. \ref{Star} and Sec. \ref{MF} has show that influence gain time scales are essentially related to a node's in-degree. Hence, one can expect different results for directed networks for which a node's in- and out-degrees are different. Our results suggest that low-in-degree out-degree hubs should be targeted independent of time horizons while in- and out-degree hubs should only be targeted when the controller's time horizon is long. Subtle differences between directed and undirected networks may motivate a more detailed investigation in future work.

It is worth noting that unlike other work in the area on competitive diffusion \citep{Goyal:2014} we have treated the opposing party as passive (or randomly) allocated in our study. Some of our results, however, suggest that for certain initial network configurations optimal influencers should shadow passively allocated influence, which is not the case for long time horizons. These findings suggest that time scales could also be crucial in competitive settings and thus a game theoretic study of competitive influence maximization with players who act on the same or different short and intermediate time horizons should be of interest for future work.

\section*{Acknowledgments}

The authors acknowledge the use of the IRIDIS High Performance Computing Facility, and associated support services at the University of Southampton, in the completion of this work. This research was sponsored by the U.S. Army Research Laboratory and the U.K. Ministry of Defence under Agreement Number W911NF-16-3-0001. The views and conclusions contained in this document are those of the authors and should not be interpreted as representing the official policies, either expressed or implied, of the U.S. Army Research Laboratory, the U.S. Government, the U.K. Ministry of Defence or the U.K. Government. The U.S. and U.K. Governments are authorized to reproduce and distribute reprints for Government purposes notwithstanding any copyright notation hereon. V.R is also partially supported by EPSRC (EP/P010164/1).

\bibliographystyle{comnet}
\bibliography{sample}

\section*{Appendix A}
\label{App:A}
In this section we show in more detail the derivations of Eq. (\ref{E3}) and Eq. (\ref{E4}). To obtain expressions (\ref{E3}) and (\ref{E4}), we solve the non-homogeneous autonomous system (\ref{E2}) by finding the general solution first, and then the particular solutions matched to the initial conditions. Equation (\ref{E2}) can be rewritten as:
\begin{equation}
\label{EA1}
\dot{\vec{x}}(t) = \vec{g} + \textbf{A}\vec{x}(t)
\end{equation}
Let us now define the \emph{fundamental matrix}, i.e., the matrix that solves the corresponding homogeneous system, as:
\begin{equation}
\boldsymbol{\varphi} = 
\left[
  \begin{array}{cc}
    \vrule & \vrule \\
    e^{\lambda_1}\vec{\xi}_1 & e^{\lambda_2}\vec{\xi}_2 \\
    \vrule & \vrule  
  \end{array}
\right]
\end{equation}
where ($\lambda_i$) and ($\vec{\xi}_i$) represent the eigenvalues and the eigenvectors of the system, respectively. Let us also define $\vec{c}$ as the vector of constants. Then,

\begin{equation}
\label{EA2}
\vec{x}(t) = \boldsymbol{\varphi}(t)\int\boldsymbol{\varphi}^{-1}(t)\vec{g}dt + \boldsymbol{\varphi}(t)\vec{c}
\end{equation}

is the general solution of the system. For the system described by Eq. (\ref{E2}), recalling that the eigenvalues are represented in Eq. (\ref{eigenvalues}) and that the eigenvectors are:
\begin{align*}
\boldsymbol{\xi}_1
 =		   \begin{bmatrix}
           -\sqrt{\frac{k}{1+k}}\\ 1
          \end{bmatrix}, 
\boldsymbol{\xi}_2 =
          \begin{bmatrix}
           \sqrt{\frac{k}{1+k}}\\ 1
          \end{bmatrix},
\end{align*}

we can derive the fundamental matrix, which is

\begin{align*}
\boldsymbol{\varphi}
 =		  \begin{bmatrix}
          -e^{(-1 - \sqrt{\frac{k}{k+1}})t}\sqrt{\frac{k}{1+k}} && 		  e^{(-1 + \sqrt{\frac{k}{k+1}})t}\sqrt{\frac{k}{1+k}} \\
          e^{(-1 - \sqrt{\frac{k}{k+1}})t} &&				               e^{(-1 + \sqrt{\frac{k}{k+1}})t} 
          \end{bmatrix},  
\end{align*}

By integrating over t the product of the inverse of $\boldsymbol{\phi}$ and the non-homogenous part $\vec{g}$, we can find the value of the integral in Eq. (\ref{EA2}), which is:
\begin{align*}
\int\boldsymbol{\varphi}^{-1}(t)\vec{g}dt
 =		  \begin{bmatrix}
          -\frac{e^{(t + \sqrt{\frac{k}{k+1}})t}}{2(k+\sqrt{k(1+k)})} \\ 		  			  e^{(t - \sqrt{\frac{k}{k+1}})t}\frac{k+\sqrt{k(1+k)}}{2k} 
          \end{bmatrix},  
\end{align*}

Now, to find the values of $\vec{c}$, we solve the system described by Eq. (\ref{EA2}) for the initial conditions $x_0(0)=x_1(0)=0$, and obtain

\begin{align*}
\vec{c}
 =		  \begin{bmatrix}
          -\frac{1+2k}{4k} \\ 		  			  \frac{1-2\sqrt{k(1+k)}}{4k} 
          \end{bmatrix},  
\end{align*}

By substituting the so obtained values of $\vec{c}$, $\int\boldsymbol{\varphi}^{-1}(t)\vec{g}dt$ and $\boldsymbol{\varphi}$ in Eq. (\ref{EA2}), and some further algebra, we obtain:

\begin{equation*}
 \label{EA3}
 x_0(t)=1+\frac{1}{2}e^{-\left(1+\sqrt{\frac{k}{1+k}}\right)t} \left(\sqrt{\frac{k}{1+k}}-1\right)
 -\frac{1}{2} e^{\left(-1+\sqrt{\frac{k}{1+k}}\right)t} \left(1+\sqrt{\frac{k}{1+k}}\right)
\end{equation*}
\begin{equation*}
\label{EA4}
 x_1(t)=1+\frac{1}{2}e^{-\left(1+\sqrt{\frac{k}{1+k}}\right)t}\left(\sqrt{\frac{1+k}{k}}-\frac{1}{k}\right)-\frac{1}{2}e^{\left(-1+\sqrt{\frac{k}{1+k}}\right)t}\left(1+\sqrt{\frac{1+k}{k}}\right)
\end{equation*} 
\raggedleft \ensuremath{\square}
%








\end{document}